\RequirePackage{fix-cm}

\documentclass[twocolumn]{svjour3}                     
\smartqed  
\usepackage{graphicx}
\usepackage{natbib}
\usepackage{amsmath}
\usepackage{amssymb}
\usepackage{subfigmat}
\usepackage{subfigure}
\usepackage{mathptmx,mathtools}
\usepackage[cmbraces,varbb,varvw]{newtxmath}
\usepackage[scaled=1.0]{helvet}
\usepackage{rotating}
\usepackage{flushend}
\usepackage{booktabs}
\usepackage{setspace}
\usepackage{color}
\usepackage{xcolor}
\usepackage{tcolorbox}
\usepackage[misc,geometry]{ifsym}
\usepackage[bookmarks = {false}]{hyperref}
\hypersetup{colorlinks = true, citecolor  = blue, urlcolor = blue, linkcolor  = blue, pdfstartview={XYZ null null 1.00}}

\usepackage{setspace}
\def\dashedrule#1#2#3{{%
\dimen1=#2 \divide\dimen1 by 2
	\def\@ruledash{%
	\rule{\dimen1}{0pt}%
	\rule[0.5ex]{#1}{2pt}%
	\rule{\dimen1}{0pt}}%

\count1=0
\loop%
\ifnum\count1<#3%
	\advance\count1 by 1%
	\@ruledash%
\repeat}}

\definecolor{light-gray}{gray}{0.75}
\begin{document}
\sloppy

\title{\vspace{-2.21cm}\begin{tcolorbox}[width=9cm, height=0.7cm, boxrule=0pt, sharp corners=all, colback=light-gray]\normalsize{\fontfamily{phv}\selectfont\normalfont\hspace{-12pt} \centering R E S E A R C H \hspace{5pt}  A R T I C L E} \end{tcolorbox} \Large Regimes of spray formation in gas-centered swirl coaxial \\ atomizers}


\titlerunning{Exp Fluids (2011) 51:587–596}        

\author{D. Sivakumar \and V. Kulkarni 
}


\institute{D. Sivakumar (\Letter)
	\and
	V. Kulkarni
	\at
	Department of Aerospace Engineering\\ 
	Indian Institute of Science \\
	Bangalore, Karnataka, 560012\\
  \email{dskumar@aero.ac.in} \\
  This is an article preprint.
}
\date{Received: 27 February 2010 / Revised: 11 March 2011 / Accepted: 11 March 2011 / Published online: 29 March 2011 \\ \scriptsize{\textcopyright} \footnotesize Springer-Verlag }


\maketitle




\begin{abstract}
Spray formation in ambient atmosphere from gas-centered swirl coaxial atomizers is described by carrying out experiments in a spray test facility. The atomizer discharges a circular air jet and an axisymmetric swirling water sheet from its coaxially arranged inner and outer orifices. A high-speed digital imaging system along with a backlight illumination arrangement is employed to record
the details of liquid sheet breakup and spray development. Spray regimes exhibiting different sheet breakup mechanisms are identified and their characteristic features presented. The identified spray regimes are wave-assisted sheet breakup, perforated sheet breakup, segmented sheet breakup, and pulsation spray regime. In the regime of  wave-assisted sheet breakup, the sheet breakup shows features similar to the breakup of two-dimensional planar air-blasted liquid sheets. At high air-to-liquid momentum
ratios, the interaction process between the axisymmetric swirling liquid sheet and the circular air jet develops spray processes which are more specific to the atomizer studied here. The spray exhibits a periodic ejection of liquid masses whose features are dominantly controlled by the central air jet.
\end{abstract}


\section{Introduction}\label{intro}
Gas-centered swirl coaxial atomizers have been employed in liquid propellant rocket engines for the purpose of fuel atomization \citep{Cohn2003, Soller2005}. The atomizer discharges an annular swirling liquid sheet and a 
\\
\\
central gas jet from the coaxially arranged outer and inner orifices, respectively. The phenomena of liquid sheet breakup and spray formation occur by means of an interaction process between the annular liquid sheet and the central gas jet. Knowledge of the breakup process of liquid sheet by the central gas jet is the key to understanding the liquid atomization or spray formation process in the atomizer. The breakup of a liquid sheet by a gas jet is extensively studied in the literature in the context of airblast atomizers used in aircraft engines \citep{Lefebvre1989}. The most studied case is the breakup of a two-dimensional planar liquid sheet by coflowing gas streams \citep{Mansour1990, Stapper1992, Lozano2001, Lozano2005, Carvalho2002, Park2004}. However, the physical configuration and flow conditions of the interacting jets in gas-centered swirl coaxial atomizers are known to influence the fundamental physical mechanisms of liquid sheet breakup, thereby altering the mechanism of spray formation. The studies on the breakup of liquid sheets began with the seminal work of \citep{Savart1833}who analyzed a flat axisymmetric sheet created by a disc-shaped obstruction in the path of a cylindrical liquid jet. The topic gained more attention after the early works on the hydrodynamic instability of liquid sheet by several prominent contributors \citep{Squire1953, Dombrowski1954, Hagerty1955, Taylor1959a, Taylor1959b, Dombrowski1963}. The study by \citep{Mansour1990} towards the understanding of air-blast atomization revealed that the presence of sinuous waves and their growth dominantly aid the breakup of a liquid sheet. At high air speed, the liquid sheet oscillates intensely and the frequency of sheet oscillation varies with the flow conditions and thickness of the liquid sheet \citep{Mansour1990, Lozano2001, Lozano2005, Carvalho2002}. The experimental study of air-blasted sheet breakup by \citep{Stapper1992} identified two different breakup mechanisms involving the formation of thin film membranes of liquid stretched between streamwise and spanwise vortical waves. Cellular type sheet breakup occurs at high air-to-liquid velocities and shows the presence of both streamwise and spanwise vortical waves which result in the formation of cell-like structures. The second mechanism, referred to as the stretched streamwise ligament breakup, occurs at low liquid velocities in which streamwise vortical waves dominate the process and result in the production of streamwise ligaments. The role of an inner gas stream on the stability of an axisymmetric liquid sheet was studied in detail \citep{Kendall1986, Lee1986,Carvalho1998, Adzic2001,Wahono2008}. In the flow regime of very low inner gas velocity, the annular liquid sheet disintegrates as spherical bubbles in a periodic manner depending on the pressure difference across the liquid sheet. \citep{Adzic2001} identified three major sheet breakup regimes: Kelvin–Helmholtz regime, cellular regime and atomization regime. As reported in a few earlier studies  \citep{Camatte1993, Lee1991}, it was observed that the inner air flow is more effective for the sheet breakup than the outer air flow. The sheet breakup process in a gas-centered swirl coaxial atomizer differs from the earlier works of axisymmetric liquid sheet breakup \citep{Kendall1986, Carvalho1998, Adzic2001} due to the presence of swirl in the liquid sheet. The presence of swirl modifies the interaction process between the liquid sheet and the central gas jet, thereby altering the mechanisms of liquid sheet breakup. The current work is a continuation of our previous research study \citep{Kulkarni2010} on the breakup of liquid sheets discharging from the gas-centered swirl coaxial atomizers. The earlier study \citep{Kulkarni2010} examined the breakup process of an outer swirling liquid sheet at different flow conditions of the central air jet. The presence of air jet at the core of the outer liquid sheet makes the conical sheet bend towards the spray axis due to the air jet driven entrainment process in the region between the inner surface of liquid sheet and the air jet. The interaction between the liquid sheet and the air jet becomes more effective as the liquid sheet draws closer to the spray axis. With increasing air jet velocity, the liquid sheet continuously shrinks along with a decrease in the sheet breakup length. The sheet atomizes into a spray of droplets in the very near region of the orifice exit at very high air jet velocity. The present study investigates breakup regimes of the outer swirling liquid sheet under the presence of the coaxially flowing central air jet. The various breakup regimes were described via a systematic analysis of high-speed video recordings of sprays discharging from the gas-centered swirl coaxial atomizers. The next section describes the atomizer configuration and the details of the experimental apparatus employed in the present work. The experimental results are discussed in Sect. \ref{resdis} and the major conclusions drawn from the work are summarized in Sect. \ref{concls}.
\vspace{-1pt}

\section{Experimental details}\label{ExpDetails}
\begin{figure}[htp!]
\vspace{-10pt}
\includegraphics[width=\linewidth]{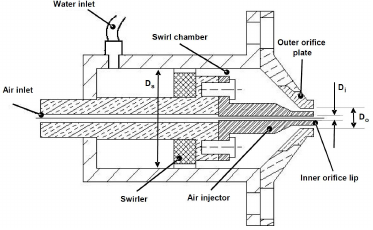}
\centering
\caption{\label{Fig1} A schematic of the gas-centered swirl coaxial atomizer used in the study. Adopted from \citep{Kulkarni2010}.}
\vspace{-10pt}
\end{figure}

A schematic of the gas-centered swirl coaxial atomizer used in the study is shown in Fig. \ref{Fig1}. Major components of the atomizer assembly were a central gas injector, an outer orifice plate, a swirler, and a swirl chamber. The central gas injector comprises a circular orifice, fabricated using electrodischarge machining (EDM) process, with internal diameter, $D_i$ = 2.4 mm and length, $L_i$ = 21.6 mm. The thickness of orifice lip, $d$ was kept at 0.3 mm. The outer orifice plate consisted of a converging portion and an orifice. The diameter, $D_o$ and length, $L_o$ of the outer orifice were kept at 4.4 and 3 mm, respectively. 

Swirling motion to the flowing liquid was imparted by passing the liquid through six starts helical passages of rectangular cross-section present over the periphery of the swirler. Two gas-centered swirl coaxial atomizer configurations, CA1 and CA2, were studied with swirlers of different geometrical parameters. The swirl number, $S$, expressed as in Kulkarni et al. (2010), for the atomizers CA1 and CA2, were 25.7 and 12.3, respectively. The diameter of swirl chamber, $D_s$ was kept at 21.4 mm. Water with density, $\rho_l =$ 998 kgm\textsuperscript{-3}, dynamic viscosity, $\mu_l$ = 1.003 $\times$ 10\textsuperscript{-3} kgm\textsuperscript{-1} s\textsuperscript{-1}, surface tension, $\sigma$ = 0.0728 Nm\textsuperscript{-1} and air with density, $\rho_g =$ 1.2 kgm\textsuperscript{-3}, dynamic viscosity, $\mu_g$ = 1.81 $\times$ 10\textsuperscript{-5} kgm\textsuperscript{-1}s\textsuperscript{-1} were used as experimental fluids. Spray experiments were carried out in a standard spray test facility. Additional details of the test facility were described in \cite{Kulkarni2010}. The mass flow rate of air, mg discharging from the central orifice of the gascentered swirl coaxial atomizer was measured using an orifice meter.

The thickness of liquid sheet is the flow geometrical scale to describe the breakup process of liquid sheets. It was estimated for different flow conditions using an analytical relation reported in the literature for swirling liquid sheets discharging from a simple pressure swirl atomizer \citep{Lefebvre1989}. The analytical relation provides the thickness of swirling liquid sheet at the orifice exit, $t$ and is expressed as,
\begin{equation}
t_f = 3.66\left(\dfrac{D_o m_l\mu_l}{\rho_l (\Delta P_l)}\right)^{0.25}
\end{equation}
where $m_l$ is the liquid mass flow rate, and $\Delta P_l$, the injection pressure drop of liquid across the atomizer. For a given $\Delta P_l$, the value of $m_l$ was estimated by measuring the volume of liquid collected in a time interval. The axial velocity of liquid sheet at the orifice exit, $U_l$ was estimated from the mass conservation as
\begin{equation}
U_l = \dfrac{m_l}{\rho_l \pi t(D_o - t)}
\end{equation}
The axial velocity of the central air jet at the orifice exit, $U_g$ was estimated from the mass conservation as
\begin{equation}
U_a = \dfrac{4m_g}{\rho_g\pi D_i^2}
\end{equation}
\begin{table}[htp!]
\caption{\label{Table1}Details of experimental conditions}
\small
\addtolength{\tabcolsep}{0em}
\begin{tabular*}{\linewidth}{@{\extracolsep{\fill}}ll@{\extracolsep{\fill}}}
\toprule
\textrm{Flow parameters} & \textrm{Values} \rule{0pt}{8pt}\\
\midrule
\textrm{$m_l$ (gs\textsuperscript{-1})}  & \textrm{23–47}\\ [2pt]
\textrm{$m_g$ (gs\textsuperscript{-1})}  & \textrm{0–1.36}\\ [2pt]
\textrm{$U_l$ (ms\textsuperscript{-1})}  & \textrm{0–1.36}\\ [2pt]
\textrm{$U_g$ (ms\textsuperscript{-1})}  & \textrm{0–250}\\ [2pt]
\textrm{$J$}  & \textrm{0–15}\\ [2pt]
\bottomrule
\end{tabular*}
\end{table}
Spray experiments were conducted for different combinations of outer liquid sheet and central gas jet flow conditions. Table \ref{Table1} shows the details of experimental conditions examined in the present study. The salient details of liquid sheet breakup were deduced from the captured images of liquid sheets. A Nikon D1X digital camera with a diffused strobe lamp system was used
to take still photographs of liquid sheets. The flash duration of the strobe lamp was approximately 15 $\mu$s which was the imaging time for the still photographs. The pixel resolution of the camera was 2000 $\times$ 1312. These still photographs were used to extract the measurements of the breakup length of liquid sheet, defined as the distance measured along the spray axis from the orifice exit to the location at which the sheet breakup first occurs. For a given flow condition, a minimum of six photographs captured at different instants were used to deduce the mean breakup length of liquid sheet, $L_b$. The temporal features of sheet breakup process were characterized by taking high-speed motion pictures of the sheet breakup. For this purpose, a high-speed camera system (Redlake Y4L) along with a backlighting source was operated with different combinations of camera frame speed and image resolution. The digital image recordings were analyzed to elucidate various features of sheet breakup.

\section{Results and discussion}\label{resdis}
In gas-centered swirl coaxial atomizers, the fluid dynamic interaction between the outer swirling liquid sheet and the central gas jet causes the breakup of the liquid sheet and subsequently the spray formation. The interaction process is similar to the one observed in the breakup of air-blasted planar liquid sheets (\citet{Mansour1990, Stapper1992, Lozano2001, Lozano2005, Carvalho2002}). However, in the present work, the geometrical configuration of the interacting jets (the axisymmetric form of the liquid sheet and the circular shape of the central air jet) can develop sheet breakup and spray formation mechanisms
which could be different from those observed with the airblasted planar liquid sheets.

\subsection{Regimes of spray formation}
\begin{figure}[htp!]
\vspace{-7pt}
\includegraphics[width=\columnwidth]{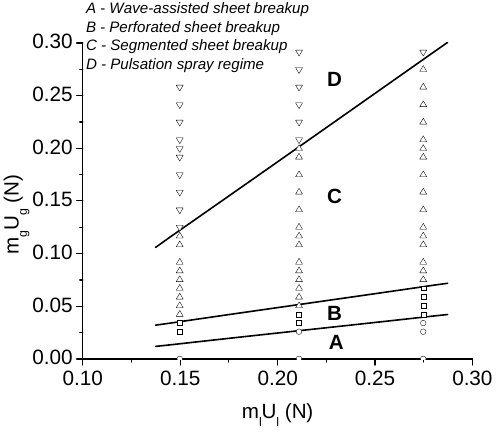}
\centering
\caption{\label{Fig2} Flow regimes of liquid sheet breakup and atomization for sprays discharging from the gas-centered swirl coaxial atomizer CA1. The symbols identify the respective regime of spray formation. Circle-wave-assisted sheet breakup, square-perforated sheet breakup, triangle—pulled-out segmented sheet breakup, and inverted triangle-pulsation spray regime}
\end{figure}
\begin{figure*}[t!]
\includegraphics[width=\linewidth]{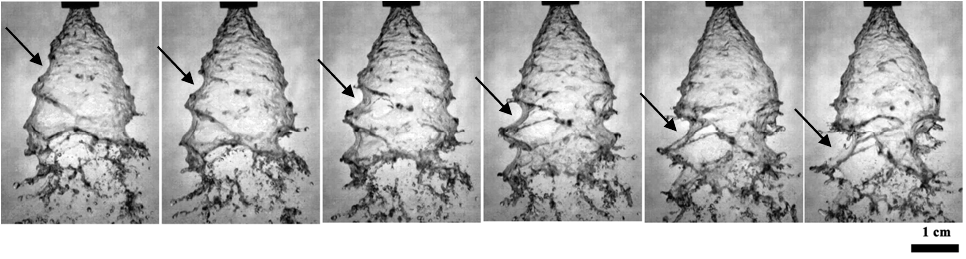}
\centering
\caption{\label{Fig3} A typical high-speed image sequence illustrating the wave assisted sheet breakup of the liquid sheet discharging from the gascentered swirl coaxial atomizer CA1. $U_l$ = 5.8 ms\textsuperscript{-1}  and $U_g$ = 77.3 ms\textsuperscript{-1}. The images are separated by a time interval of 0.714 ms. The \textit{arrow} marks point towards a growing sinuous wave}
\end{figure*}
A detailed analysis of high-speed motion pictures of liquid sheet breakup and atomization for sprays discharging from the gas-centered swirl coaxial atomizer at different combinations of $U_l$ and $U_g$ paved the way to identify the various regimes of spray formation process from the atomizer. For a given atomizer, the analysis examined the cases of three liquid flow conditions in combination with several tens of central air flow conditions. The typical outcome of this analysis is highlighted in Fig. \ref{Fig2} which shows the variation of $m_lU_l$ (liquid momentum) versus $m_gU_g$ (central air jet momentum). For each $U_l$, the liquid sheet exhibits different mechanisms of sheet breakup and spray formation depending on the flow condition of central air jet.

The analysis identified four different regimes: wave-assisted sheet breakup, perforated sheet breakup, segmented sheet breakup, and pulsation spray regime. With increasing $U_g$, the nature of fluid dynamic interaction process between the liquid sheet and the central air jet gets changed and hence the mechanism of spray formation. In the wave-assisted and perforated sheet breakups, an intact axisymmetric liquid sheet exists for all flow conditions. 

Such a feature is absent in the last two regimes where the liquid mass is fragmented into ligaments and droplets in the very near region of the orifice exit. It must be mentioned here that the transition line separating the wave-assisted sheet breakup and the perforated sheet breakup is based on individual judgment. This means that wave features on the liquid sheet may be seen for the flow conditions marked in the perforated sheet breakup regime in Fig. \ref{Fig2} and vice versa. A systematic analysis of the salient features of these spray regimes is given in the ensuing sections.

\subsection{Wave-assisted and perforated sheet breakups}
\subsubsection{Wave characteristics}

The wave-assisted sheet breakup occurs via the growth of waves on the liquid sheet. Previous studies \citep{Mansour1990, Lozano2001} on the breakup of airblasted liquid sheets confirmed that an efficient sheet breakup is caused by the growth of sinuous waves. The breakup of liquid sheet occurs when the wave amplitude reaches a critical value \citep{Dombrowski1963}. Figure \ref{Fig3} shows a typical high-speed image sequence illustrating the wave-assisted sheet breakup of the liquid sheet with $U_l$ = 5.8 ms\textsuperscript{-1} discharging from the atomizer. It is observed from the image sequence that surface waves developed near the orifice exit grow rapidly in the downstream and tear out at the breakup location. The high-speed image recordings were used to deduce the wavelength, $k$ and wave propagation velocity, $c$ of the surface waves. 
\begin{figure}
\includegraphics[scale =0.9]{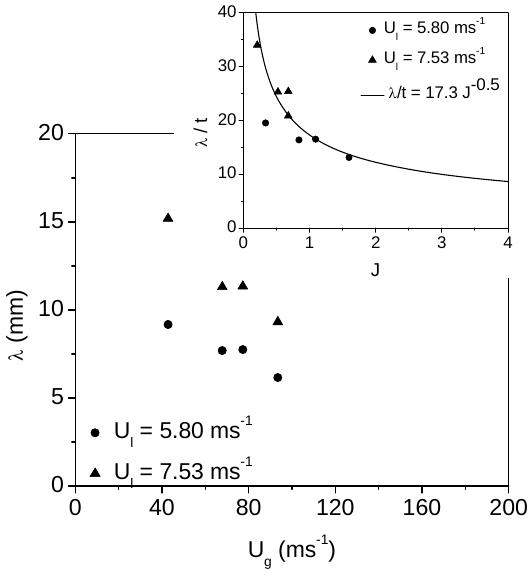}
\centering
\caption{\label{Fig4} The variation of $k$ with $U_g$ for different liquid sheets discharging from the gas-centered swirl coaxial atomizer CA1. The \textit{insert} shows the variation of nondimensionalized wavelength, $\lambda/t$ with $J$}
\vspace{-11pt}
\end{figure}
At a given flow condition, more than five sets of images featuring sinuous waves were used to deduce the averaged measurements of $k$ and $c$. Figure \ref{Fig4} shows the variation of $k$ with $U_g$ for different liquid sheets discharging from the atomizer. A variation of $k$ along the axial direction was noticed during the extraction of measurements from the images. The measured values of $k$ given in Fig. \ref{Fig4} correspond to the wave segments seen immediate upstream of the breakup location. The wavelength decreases with increasing $U_g$ and increases with increasing $U_l$. These trends are in agreement with those reported for the breakup of air-blasted liquid sheets \citep{Lozano2001,Lozano2005, Carvalho2002}. The insert of the figure shows the variation of nondimensionalized wavelength, $\lambda/t$ with air-to-liquid momentum ratio, $J$, defined here as $\rho_gU_g^2 D_i/\rho_lU_l^2 t$, for different $U_l$ and the continuous line traces the variation of $\lambda/t = $ 17.3 $J^{-0.5}$ in accordance with the trend $\lambda/t \propto J^{-0.5}$ reported by \citep{Lozano2005} for the air-blasted liquid sheets. 

\begin{figure}[htp!]
\vspace{-10pt}
\includegraphics[scale =0.9]{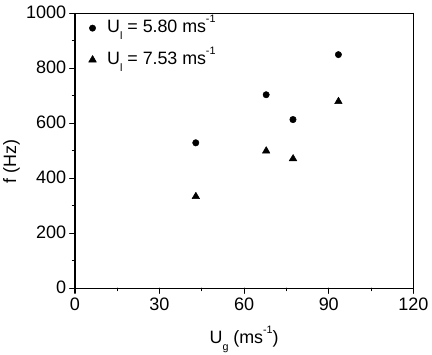}
\centering
\caption{\label{Fig5} The variation of wave frequency, $f$ with $U_g$ for different liquid sheets discharging from the gas-centered swirl coaxial atomizer CA1}
\end{figure}

The measurements of $c$ obtained from the high-speed image recordings of liquid sheet breakup along with the measurements of $k$ pave the way for the estimation of the wave frequency, $f$ as $f = c/k$. Figure \ref{Fig5} shows the variation of $f$ with $U_g$ for different liquid sheets discharging from the gas-centered swirl coaxial atomizer. The wave frequency increases with increasing $U_g$ which highlights the significant influence of air jet characteristics on the sheet breakup process. The wave frequency is lesser for the higher inertia liquid sheet for a given $U_g$ because larger inertia takes the swirling liquid sheet farther from the spray axis, thereby delaying effective interaction between the liquid sheet and the air jet.
\vspace{-10pt}
\subsubsection{Perforated sheet breakup}
\begin{figure}[t!]
\includegraphics[scale=0.9]{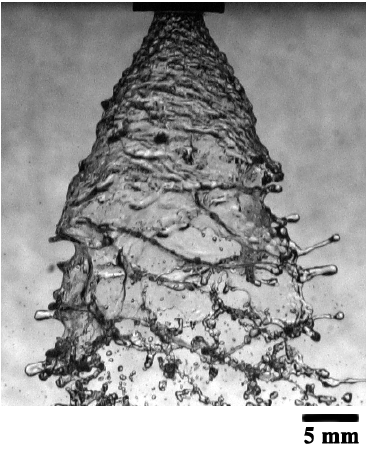}
\centering
\caption{\label{Fig6} A typical photograph illustrating perforated sheet breakup for the liquid sheet with $U_l =$ 6.38 ms\textsuperscript{-1} and $U_g =$ 117.4 ms\textsuperscript{-1} discharging from the gas-centered swirl coaxial atomizer CA1.}
\vspace{-15pt}
\end{figure}

The dominance of wave features on the liquid sheet gradually disappears as $U_g$ nears the dashed line separating the wave-assisted sheet breakup and the perforated sheet breakup given in Fig. \ref{Fig2}. The increasing spray contraction with increasing $U_g$ enhances the interaction process between the liquid sheet and the air jet. In this regime, the breakup process is marked by the presence of cellular structures (thin film membranes surrounded by thick ligaments) and perforations on the liquid sheet. A typical image illustrating such features on the surface of liquid sheet discharging from the gas-centered swirl coaxial atomizer is given in Fig. \ref{Fig6}.
\vspace{-5pt}

\subsubsection{Sheet breakup length}
\begin{figure}[b!]
\vspace{-2pt}
\includegraphics[width=\columnwidth]{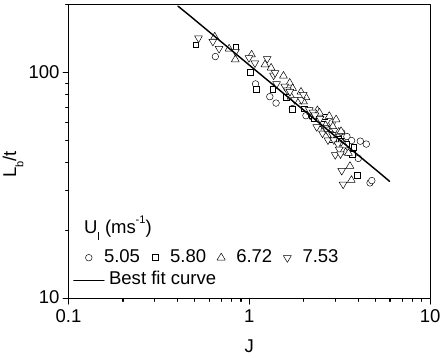}
\centering
\caption{\label{Fig7}The variation of nondimensionalized $L_b$ with $J$ for different liquid sheets discharging from the gas-centered swirl coaxial atomizer CA1}
\vspace{0pt}
\end{figure}
\vspace{0pt}
\begin{figure*}[t!]
\includegraphics[width=\textwidth]{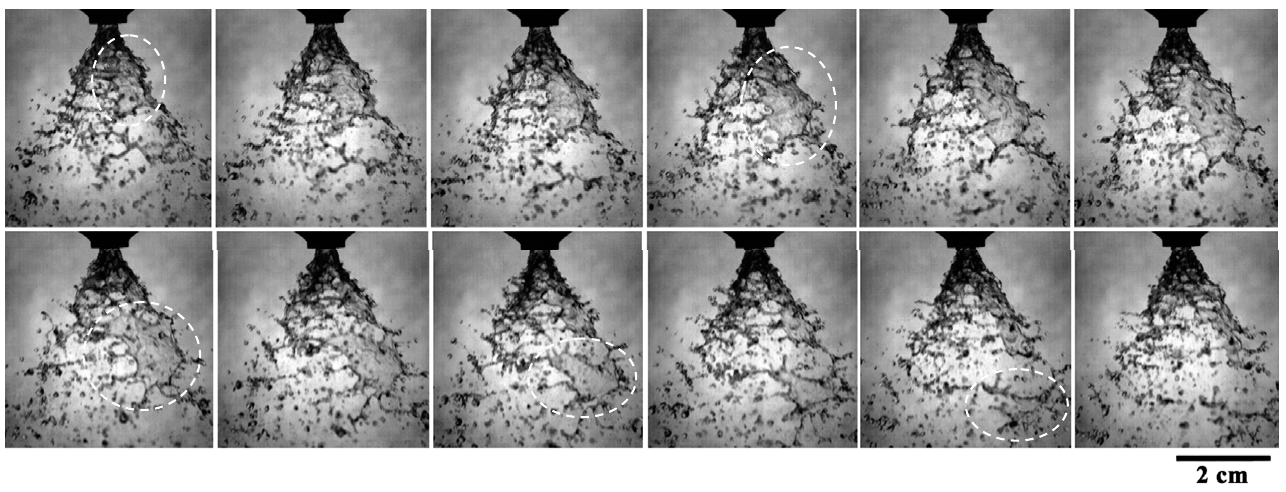}
\centering
\caption{\label{Fig8} A high-speed image sequence illustrating the pulled-out segmented sheet breakup observed during the spray formation in the gascentered swirl coaxial atomizer. U\textsubscript{\textit{l}} = 5.8 ms\textsuperscript{-1} and U\textsubscript{\textit{g}} = 147 ms\textsuperscript{-1}. The images are separated by a time interval of 0.714 ms.}
\end{figure*}
\vspace{-3pt}
\citet{Kulkarni2010} described the quantitative variation of $L_b$ with $U_g$ for different flow conditions of the outer liquid sheet. For lower values of $U_g$, $L_b$ remains constant with $U_g$ due to negligible interaction between the liquid sheet and the air jet. An effective interaction between the jets can be realized only after the liquid sheet bends towards the spray axis. For higher values of $U_g$, the liquid sheet shows the influence of air jet as the sheet breakup length decreases steadily with increasing $U_g$ \citet{Kulkarni2010}. The significant role of the central air jet on the variation of $L_b$ suggests that the sheet breakup is governed by the transfer of energy from the air jet to the liquid sheet similar to the one seen with the breakup of air-blasted liquid sheets \citet{Mansour1990, Stapper1992, Lozano2001, Lozano2005, Carvalho2002}. An analysis of the present measurements of $L_b$ in accordance with the earlier studies on air-blasted liquid sheets \citet{Carvalho2002} was carried out. In this analysis, the quantitative measurements of $L_b$ were nondimensionalized with the thickness of liquid sheet at the orifice exit, $t$, an appropriate length scale for the present problem. The flow conditions of interacting jets were represented in terms of $J$. Figure \ref{Fig7} shows the variation of nondimensionalized $L_b$ with $J$ for different liquid sheets discharging from the atomizer. Note that the measurements of $L_b$ for the lower values of $U_g$ at which the air jet does not influence the sheet breakup were carefully eliminated from the analysis. The measurements of $L_b$ collapse into a single variation satisfactorily as seen in the figure. The best fit curve for the measurements, highlighted as thick black line in Fig. \ref{Fig7}, exhibits the variation $L_b/t \propto J^{-0.663}$. The index value $-$0.663 is very close to the index value $-$0.68 obtained by \citet{Carvalho2002} for the air-blasted liquid sheets. This analysis along with the previously discussed wave characteristics of the liquid sheet suggests that the waveassisted sheet breakup mechanism of liquid sheets in gascentered swirl coaxial atomizer may be similar to the mechanism seen in the breakup of air-blasted liquid sheets.
\vspace{-9pt}
\subsection{Segmented sheet breakup}
For a given liquid sheet condition, the segmented sheet breakup occurs in relatively higher values of $U_g$. The axisymmetric
form of liquid sheet breakup, as seen in the wave-assisted and perforated sheet breakup processes, is no longer seen in this regime of sheet breakup. The high velocity air jet pulls the liquid sheet very near to the spray axis and impinges directly on the inner surface of the liquid sheet and the spray formation occurs in the vicinity of the orifice exit. Figure \ref{Fig8} shows a high-speed image sequence illustrating the process of spray formation under this breakup regime. At first, a segmented portion of liquid sheet, identified in the figure within the region enclosed by dashed lines, pulls itself away from the spray axis by the centrifugal force of the liquid mass and enlarges with time. At some instant, the segmented liquid sheet detaches itself from the rest of liquid mass and thereafter gradually collapses into drops. The analysis of high-speed image recordings revealed that this type of sheet breakup occurs continuously, however, with random time intervals in a certain range of flow conditions depending on the values of $U_l$ and $U_g$. 
\begin{figure*}[t!]
\includegraphics[scale=0.9]{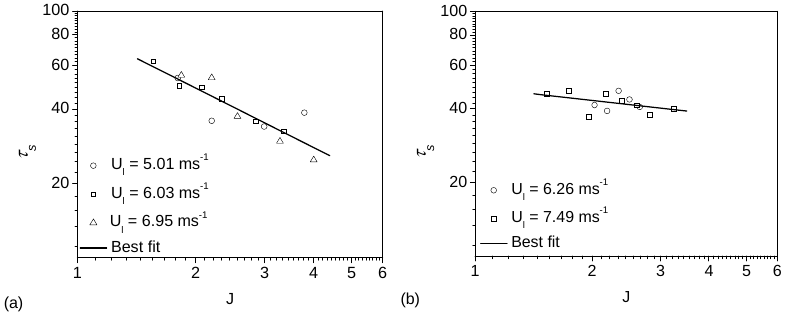}
\centering
\caption{\label{Fig9}The variation of nondimensionalized collapsing time, $\tau_s$ of the segmented liquid sheets with $J$ for different flow conditions of outer liquid sheet discharging from the gascentered swirl coaxial atomizers. \textbf{a} CA1, and \textbf{b} CA2.}
\end{figure*}
\begin{figure*}[htp!]
\includegraphics[scale=0.9]{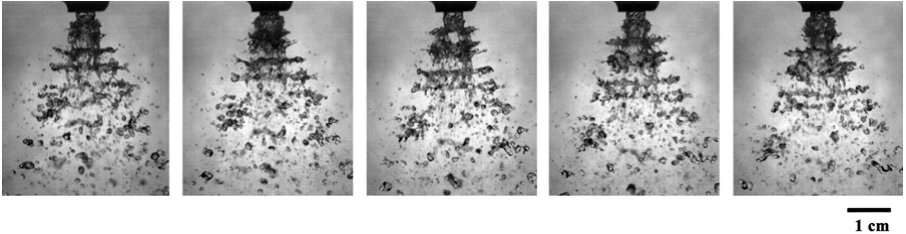}
\centering
\caption{\label{Fig10}A continuous image sequence illustrating the periodic ejection of liquid masses in sprays discharging from the gas-centered swirl coaxial atomizer CA1. $U_l = $ 3.74 ms\textsuperscript{-1} and $U_l =$ 147 ms\textsuperscript{-1}. The images are separated by a time interval of 0.714 ms}
\end{figure*}
\begin{figure*}[htp!]
\includegraphics[scale=0.9]{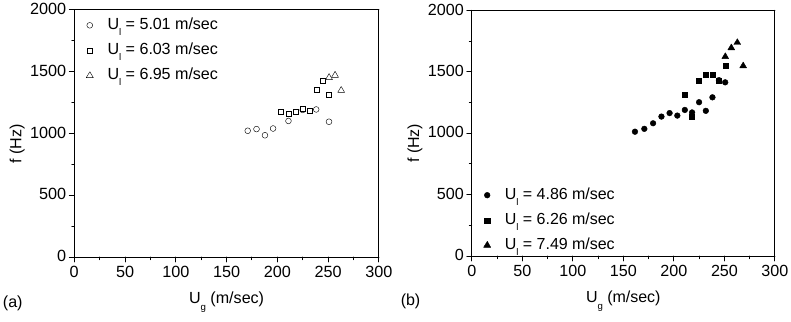}
\centering
\caption{\label{Fig11} The variation of spray (b) pulsation frequency, \textit{f}\textsubscript{p} with U\textsubscript{g}  for different flow conditions of sprays discharging from the gas-centered swirl coaxial atomizers. a CA1, and b CA2}
\end{figure*}
The image recordings enabled estimation of the collapsing time, $\tau_s$ of the segmented sheet, defined as the time lapsed from the identification of a segmented sheet to its complete collapse into drops, for different conditions of $U_l$ and $U_g$. For a particular combination of $U_l$ and $U_g$, the value of $\tau_s$ was obtained by averaging the collapsing time of several segmented sheets originating at different instants. Figure \ref{Fig9} shows the variation of nondimensionalized collapsing time, $\tau_s = T_s U_l/t$ with $J$ for liquid sheets discharging from the gas-centered swirl coaxial atomizers. The trend suggests that the breakup and collapse processes of the segmented liquid sheets accelerate with increasing \textit{J}. A comparison of measurements given in Fig. \ref{Fig9} \textcolor{blue}{a} and \textcolor{blue}{b} shows that the segmenting and collapsing processes are influenced by the swirling intensity of the liquid sheets. In this regime, the spray formation process is more complicated than that seen in the regimes discussed in the previous sections. The sheet breakup process highlighted in Fig. \ref{Fig8} is more clearly seen for lower values of $U_g$. The shedding of sheet segments becomes more complex at higher values of $U_g$. The typical size of the sheet segment formed near the orifice exit decreases with increasing $U_g$. The available experimental data suggest that the shedding of sheet segments is a random process. Approximately 600$-$800 sheet segments per second are seen in a typical test case. More experiments are needed to provide further quantitative details of sheet segmenting process. 
\vspace{-8pt}
\subsection{Pulsation spray regime}

This type of spray formation in the gas-centered swirl coaxial atomizer occurs in a flow regime where the air jet momentum is comparable to the liquid sheet momentum. The higher momentum central air jet shatters the bulk liquid masses at the orifice exit itself. The interaction and mixing processes between the liquid sheet and the central air jet result in a periodic ejection of liquid masses in the very near region of the orifice exit. Figure \ref{Fig10} shows a typical highspeed image sequence illustrating the periodic ejection of liquid masses/ligaments during the spray formation under this flow regime of the gas-centered swirl coaxial atomizer. The ejected liquid masses stretch along the radial direction as they flow in the downstream and breakup further into drops. Figure \ref{Fig11} shows the variation of spray pulsation frequency, $f_p$ with $U_g$ for different conditions of liquid sheets discharging from the gas-centered swirl coaxial atomizers CA1 and CA2. For a given $U_l$, the spray pulsation frequency increases with increasing $U_g$ as seen in Fig. \ref{Fig11}. The frequency measurements given in Fig. \ref{Fig11} are expressed in Fig. \ref{Fig12} in terms of Strouhal number, \textit{St} defined as $St = f_pD_i/U_g$. The estimated values of $St$ for the pulsation frequency measurements collapse into an almost constant value as shown in Fig. \ref{Fig12}.

The dynamics of pulsating sprays discharging from coaxial injectors have been investigated earlier \citep{Bazarov1998, Im2009} studied the pulsation characteristics of sprays from a gas$-$liquid swirl coaxial injector discharging a central swirling liquid sheet and an annular gas jet. The authors observed that the frequency of spray oscillation is linearly proportional to the liquid and gas axial Reynolds numbers. It was concluded that the spray pulsation behavior is attributed to the unstable waves of the liquid sheet. The present gas-centered swirl coaxial atomizer is different from the atomizer used by \cite{Im2009} and hence the interaction process between the liquid sheet and the air jet. The measurements described above suggest that the frequency of spray pulsation is almost independent of the flow conditions of outer liquid sheet. In the spray pulsation regime, the high velocity central air jet may be developing a closed annular liquid jet zone just at the orifice exit due to the decrease in static pressure inside the liquid sheet. The incoming air mass ruptures the closed liquid jet zone immediately, thereby creating radially spreading liquid ligaments. The rupturing process triggers the development of the closed annular liquid jet zone. This process occurs intermittently and develops the spray pulsation in the gas-centered swirl coaxial atomizer.

\subsection{Role of swirling intensity in spray formation}

As mentioned earlier, the present work of spray formation in gas-centered swirl coaxial atomizers differs from the earlier studies of air-blast type atomizers \citep{Mansour1990, Stapper1992, Carvalho1998, Adzic2001, Lozano2001, Carvalho2002} mainly due to the presence of swirling intensity. The experimental measurements suggest that the regime of spray formation is influenced by the variation in swirling intensity. For instance, the transition from segmented sheet breakup to pulsation spray regime with increasing central air jet momentum for a given liquid sheet momentum occurs at higher air jet momentum for the liquid sheets with higher swirling intensity. 
\begin{figure}[htp!]
\includegraphics[scale=0.9]{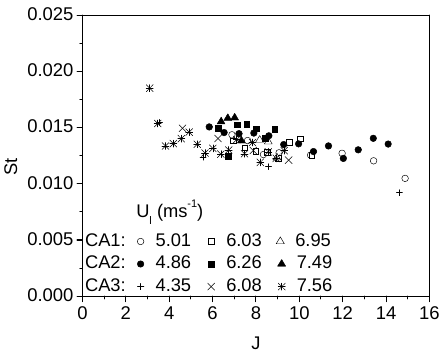}
\centering
\caption{\label{Fig12} The variation of $St$ of spray pulsation with $J$ for sprays discharging from the gas-centered swirl coaxial atomizers}
\end{figure}
This is attributed to the fact that a liquid sheet with larger spray cone angle requires higher air jet momentum to ensure effective interaction between the liquid sheet and the air jet. Nevertheless, the influence of swirling intensity on the characteristic flow features of different spray regimes is small or almost negligible. In the wave-assisted sheet breakup regime, the analysis of currently available experimental measurements shows that the nondimensionalized sheet breakup length exhibits the trend $L_b/t \propto J^{-0.663}$ and $L_b/t \propto J^{-0.649}$ for the atomizers CA1 and CA2, respectively. The variation in the index value of $J$ between the two atomizers is almost negligible. In the regime of segmented sheet breakup, as seen in Fig. \ref{Fig9}, the influence of swirling intensity is slightly seen. The plot given in Fig. \ref{Fig12} clearly shows that the spray pulsation is not influenced by swirling intensity of the liquid sheet. Thus, for the gas-centered swirl coaxial atomizers examined in the present study, the dynamics of sheet breakup and spray pulsation is dominantly governed by $J$. A detailed study involving atomizers with different geometrical parameters and swirling intensity is needed to strengthen this observation further.
\vspace{-11pt}
\section{Conclusions}\label{concls}
The process of spray formation from gas-centered swirl coaxial atomizer employed in liquid propellant rocket engines is studied with water and air as experimental fluids. The fluid dynamic interaction between the outer swirling liquid sheet and the central air jet and the subsequent spray development are characterized with the aid of visualization and photographic techniques. The characterization of spray formation using high-speed video photography revealed four different spray regimes for sprays discharging from the gas-centered swirl coaxial atomizer with different combinations of inner air and outer liquid flow conditions:
wave-assisted sheet breakup, perforated sheet breakup, segmented sheet breakup, and pulsation spray regime. In the regime of wave-assisted sheet breakup, the breakup of liquid sheet occurs via the growth of sinuous waves. For a given liquid sheet condition, the wavelength measured in the near region of sheet breakup decreases and the wave frequency, estimated from the measurements of wavelength and wave propagation velocity, increases with increasing velocity of the central air jet. The measured
sheet breakup length scales with air-to-liquid momentum ratio as observed in the breakup of two-dimensional planar air-blasted liquid sheets. The sheet breakup length decreases with increasing air-to-liquid momentum ratio. The available measurements suggest that the wave-assisted sheet breakup process of gas-centered swirl coaxial atomizer resembles the breakup behavior of two-dimensional planar air-blasted liquid sheets. The last two spray regimes are more specific to the gas-centered swirl coaxial atomizer. No intact axisymmetric liquid sheet is seen and the bulk liquid mass is fully fragmented into spray drops in these regimes. In the segmented sheet breakup, the spray development is featured by random formation and collapse of segmented sheets. For a given liquid sheet condition, the time associated with the formation and collapse of segmented sheets decreases with increasing velocity of the central air jet. At high air-to-liquid momentum ratios, the spray exhibits a periodic ejection of liquid masses whose
behavior is predominantly governed by the central air jet. The frequency of spray pulsation increases with velocity of the central air jet for a given liquid sheet condition. The estimated Strouhal number for this oscillatory spray flow is found to be in the range 0.012$-$0.015. Limited results show that the influence of swirling intensity on the characteristic features of sheet breakup and spray pulsation may be negligible.

\vspace{0pt}
\begin{acknowledgements}
This work is supported by Space Technology Cell, Indian Institute of Science under the Grant ISTC/MAE/DS/235. The authors gratefully acknowledge the funding from University Grants Commission, India for the establishment of high-speed digital imaging
facility at the laboratory. The authors would like to thank Mr. U. Satish Babu for his assistance in conducting the spray experiments.
\end{acknowledgements}
\vspace{0pt}

\vspace{-2.5pt}
\bibliographystyle{spbasic}      
\footnotesize\bibliography{Refs_Sivakumar_Kulkarni_2011}   

\begin{thebibliography}{26}
\providecommand{\natexlab}[1]{#1}
\providecommand{\url}[1]{{#1}}
\providecommand{\urlprefix}{URL }
\expandafter\ifx\csname urlstyle\endcsname\relax
  \providecommand{\doi}[1]{DOI~\discretionary{}{}{}#1}\else
  \providecommand{\doi}{DOI~\discretionary{}{}{}\begingroup
  \urlstyle{rm}\Url}\fi
\providecommand{\eprint}[2][]{\url{#2}}

\bibitem[{Adzic et~al(2001)Adzic, Carvalho, and Heitor}]{Adzic2001}
Adzic M, Carvalho I, Heitor M (2001)
  \href{http://refhub.elsevier.com/S0894-1777(15)00119-3/h0130}{Visualization
  of the disintegration of an annular liquid sheet in a coaxial air blast
  injector at low atomizing air velocities}. Optical Diagnostics in Engineering
  5(1):27--38

\bibitem[{Bazarov and Yang(1998)}]{Bazarov1998}
Bazarov VG, Yang V (1998)
  \href{https://doi.org/10.2514/2.5343}{Liquid-propellant rocket engine
  injector dynamics}. Journal of Propulsion and Power 14(5):797--806

\bibitem[{Camatte et~al(1993)Camatte, Car{\'e}, Dumouchel, and
  Ledoux}]{Camatte1993}
Camatte P, Car{\'e} I, Dumouchel C, Ledoux M (1993)
  \href{https://doi.org/10.1007/978-1-4899-1594-8_5}{Modelisation of
  pulverisation systems: some aspects of linear stability analysis}.
  Instabilities in Multiphase Flows pp 53--67

\bibitem[{Carvalho and Heitor(1998)}]{Carvalho1998}
Carvalho I, Heitor M (1998) \href{https://doi.org/10.1007/s003480050190}{Liquid
  film break-up in a model of a prefilming airblast nozzle}. Experiments in
  Fluids 24(5):408--415

\bibitem[{Carvalho et~al(2002)Carvalho, Heitor, and Santos}]{Carvalho2002}
Carvalho I, Heitor M, Santos D (2002)
  \href{https://doi.org/10.1016/S0301-9322(01)00088-X}{Liquid film
  disintegration regimes and proposed correlations}. International journal of
  multiphase flow 28(5):773--789

\bibitem[{Cohn et~al(2003)Cohn, Strakey, Bates, Talley, Muss, and
  Johnson}]{Cohn2003}
Cohn R, Strakey P, Bates R, Talley D, Muss J, Johnson C (2003) Swirl coaxial
  injector development. In: \href{https://doi.org/10.2514/6.2003-125}{41st
  Aerospace Sciences Meeting and Exhibit}, p 125

\bibitem[{Dombrowski and Fraser(1954)}]{Dombrowski1954}
Dombrowski N, Fraser RP (1954) \href{https://doi.org/10.1098/rsta.1954.0014}{A
  photographic investigation into the disintegration of liquid sheets}.
  Philosophical Transactions of the Royal Society of London Series A,
  Mathematical and Physical Sciences 247(924):101--130

\bibitem[{Dombrowski and Johns(1963)}]{Dombrowski1963}
Dombrowski N, Johns W (1963)
  \href{https://doi.org/10.1016/0009-2509(63)85005-8}{The aerodynamic
  instability and disintegration of viscous liquid sheets}. Chemical
  Engineering Science 18(3):203--214

\bibitem[{Hagerty and Shea(1955)}]{Hagerty1955}
Hagerty W, Shea J (1955) A study of the stability of moving liquid film. ASME J
  Appl Mech 22:509--514

\bibitem[{Im et~al(2009)Im, Kim, Han, Yoon, and Bazarov}]{Im2009}
Im JH, Kim D, Han P, Yoon Y, Bazarov V (2009)
  \href{https://doi.org/10.1615/AtomizSpr.v19.i1.40}{Self-pulsation
  characteristics of a gas-liquid swirl coaxial injector}. Atomization and
  Sprays 19(1)

\bibitem[{Kendall(1986)}]{Kendall1986}
Kendall JM (1986) \href{https://doi.org/10.1063/1.865595}{Experiments on
  annular liquid jet instability and on the formation of liquid shells}. The
  Physics of fluids 29(7):2086--2094

\bibitem[{Kulkarni et~al(2010)Kulkarni, Sivakumar, Oommen, and
  Tharakan}]{Kulkarni2010}
Kulkarni V, Sivakumar D, Oommen C, Tharakan T (2010)
  \href{https://doi.org/10.1115/1.4000737}{Liquid sheet breakup in gas-centered
  swirl coaxial atomizers}. Journal of Fluids Engineering 132(1):011,303

\bibitem[{Lee and Wang(1986)}]{Lee1986}
Lee C, Wang T (1986) \href{https://doi.org/10.1063/1.865594}{A theoretical
  model for the annular jet instability}. The Physics of fluids
  29(7):2076--2085

\bibitem[{Lee and Chen(1991)}]{Lee1991}
Lee JG, Chen LD (1991) \href{https://doi.org/10.2514/3.10779}{Linear stability
  analysis of gas-liquid interface}. AIAA journal 29(10):1589--1595

\bibitem[{Lefebvre(1989)}]{Lefebvre1989}
Lefebvre AH (1989) Atomization and sprays, hemisphere pub. Corp, New York 1989

\bibitem[{Lozano et~al(2001)Lozano, Barreras, Hauke, and Dopazo}]{Lozano2001}
Lozano A, Barreras F, Hauke G, Dopazo C (2001)
  \href{https://doi.org/10.1017/S0022112001004268}{Longitudinal instabilities
  in an air-blasted liquid sheet}. Journal of Fluid Mechanics 437:143--173

\bibitem[{Lozano et~al(2005)Lozano, Barreras~Toledo, Siegler, and
  L{\"o}w}]{Lozano2005}
Lozano A, Barreras~Toledo F, Siegler C, L{\"o}w D (2005)
  \href{https://doi.org/10.1007/s00348-005-0989-1}{The Effects of Sheet
  Thickness on the Oscillation of an Air-Blasted Liquid Sheet}. Experiments in
  Fluids 39:127--139

\bibitem[{Mansour and Chigier(1990)}]{Mansour1990}
Mansour A, Chigier N (1990)
  \href{https://doi.org/10.1063/1.857724}{Disintegration of liquid sheets}.
  Physics of fluids A: fluid dynamics 2(5):706--719

\bibitem[{Park et~al(2004)Park, Huh, Li, and Renksizbulut}]{Park2004}
Park J, Huh KY, Li X, Renksizbulut M (2004)
  \href{https://doi.org/10.1063/1.1644575}{Experimental investigation on
  cellular breakup of a planar liquid sheet from an air-blast nozzle}. Physics
  of Fluids 16(3):625--632

\bibitem[{Savart(1833)}]{Savart1833}
Savart F (1833) Memoir on the impact of a liquid vein launched against a
  circular plane. Ann chem 54(56):1833

\bibitem[{Soller et~al(2005)Soller, Wagner, Kau, Martin, and
  Maeding}]{Soller2005}
Soller S, Wagner R, Kau HP, Martin P, Maeding C (2005) Characterisation of main
  chamber injectors for gox/kerosene in a single element rocket combustor. In:
  \href{https://doi.org/10.2514/6.2005-3750}{41st AIAA/ASME/SAE/ASEE Joint
  Propulsion Conference \& Exhibit}, p 3750

\bibitem[{Squire(1953)}]{Squire1953}
Squire H (1953) \href{https://doi.org/10.1088/0508-3443/4/6/302}{Investigation
  of the instability of a moving liquid film}. British Journal of applied
  physics 4(6):167

\bibitem[{Stapper et~al(1992)Stapper, Sowa, and Samuelsen}]{Stapper1992}
Stapper B, Sowa W, Samuelsen G (1992)
  \href{https://doi.org/10.1115/1.2906305}{An experimental study of the effects
  of liquid properties on the breakup of a two-dimensional liquid sheet}. J Eng
  Gas Turbines Power 114(1):39--45

\bibitem[{Taylor(1959{\natexlab{a}})}]{Taylor1959a}
Taylor GI (1959{\natexlab{a}})
  \href{https://doi.org/10.1098/rspa.1959.0195}{The dynamics of thin sheets of
  fluid II. Waves on fluid sheets}. Proceedings of the Royal Society of London
  Series A Mathematical and Physical Sciences 253(1274):296--312

\bibitem[{Taylor(1959{\natexlab{b}})}]{Taylor1959b}
Taylor GI (1959{\natexlab{b}})
  \href{https://doi.org/10.1098/rspa.1959.0196}{The dynamics of thin sheets of
  fluid. III. Disintegration of fluid sheets}. Proceedings of the Royal Society
  of London Series A Mathematical and Physical Sciences 253(1274):313--321

\bibitem[{Wahono et~al(2008)Wahono, Honnery, Soria, and Ghojel}]{Wahono2008}
Wahono S, Honnery D, Soria J, Ghojel J (2008)
  \href{https://doi.org/10.1007/s00348-007-0361-8}{High-speed visualisation of
  primary break-up of an annular liquid sheet}. Experiments in fluids
  44:451--459

\end{thebibliography}





\end{document}